%% file: main.tex
\documentclass[acmtog,nonacm]{acmart}

\input{pre}

\AtBeginDocument{%
  }

\begin{document}

\title{Variational Green's Functions for Volumetric PDEs}

\author{João Teixeira}
\email{joao@cs.toronto.edu}
\affiliation{%
  \institution{University of Toronto}
  \city{Toronto}
  \state{ON}
  \country{CA}
}

\author{Eitan Grinspun}
\email{eitan@cs.toronto.edu}
\affiliation{%
  \institution{University of Toronto}
  \city{Toronto}
  \state{ON}
  \country{CA}
}

\author{Otman Benchekroun}
\email{otman@cs.toronto.edu}
\affiliation{%
  \institution{University of Toronto}
  \city{Toronto}
  \state{ON}
  \country{CA}
}

\renewcommand{\shortauthors}{Teixeira et al.}

\begin{abstract}
  Green's functions characterize the fundamental solutions of partial differential equations; they are essential for tasks ranging from shape analysis to physical simulation, yet they remain computationally prohibitive to evaluate on arbitrary geometric discretizations. We present Variational Green's Function (VGF), a method that learns a smooth, differentiable representation of the Green's function for linear self-adjoint PDE operators, including the Poisson, the screened Poisson, and the biharmonic equations.
  To resolve the sharp singularities characteristic of the Green's functions, our method decomposes the Green's function into an analytic free-space component, and a learned corrector component.
  Our method leverages a variational foundation to impose Neumann boundary conditions \textit{naturally}, and imposes Dirichlet boundary conditions via a projective layer on the output of the neural field. 
  The resulting Green's functions are fast to evaluate, differentiable with respect to source application, and can be conditioned on other signals parameterizing our geometry. 
\end{abstract}

\keywords{Green's functions, neural fields, partial differential equations.}

\begin{teaserfigure}
  \centering
    \includegraphics[width=\textwidth]{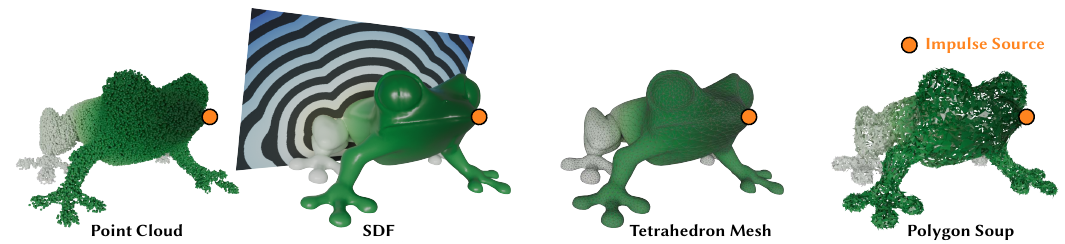}
        \captionsetup{skip=0pt}
    \caption{
    \textbf{We adopt a variational formulation for Green's functions and develop an algorithm capable of computing Green's functions on arbitrary geometric representations, such as point clouds, SDFs, tetrahedralized meshes, or polygon soups.}
  }
  \label{fig:teaser}
\end{teaserfigure}

\received{20 February 2007}
\received[revised]{12 March 2009}
\received[accepted]{5 June 2009}

\maketitle

\input{introduction}

\input{related_work_shorter}

\input{method_otman}

\input{experiments}

\input{conclusion}

\FloatBarrier
\clearpage
\bibliographystyle{ACM-Reference-Format}
\bibliography{bibliography}

\clearpage

\FloatBarrier

\input{figures}
\end{document}

%% file: pre.tex
\usepackage{xcolor}
\usepackage{colortbl}
\usepackage{cleveref}

\usepackage{wrapfig}
\usepackage{placeins}
\usepackage{makecell}

\DeclareMathOperator*{\argmin}{argmin}

\usepackage{mfirstuc}
\newcommand{\reffig}[1]{Fig.~\ref{fig:#1}}
\newcommand{\refeq}[1]{Eq.~(\ref{eq:#1})}
\newcommand{\refsec}[1]{Sec.~\ref{sec:#1}}
\newcommand{\reftab}[1]{Table.~\ref{tab:#1}}

\newcommand{\mat}[1]{\mathbf{#1}}

\newcommand{\Rn}[1]{\mathbb{R}^{#1}}

%% file: introduction.tex
\section{Introduction}
\label{sec:introduction}
\vspace{-0.3em}
\begin{quote}
``Give it a little tappy. Tap, tap, taparoo.''\\
---Happy Gilmore 
\end{quote}
\vspace{-0.4em}

\begin{figure}[b]
\includegraphics[width=1\columnwidth,keepaspectratio]{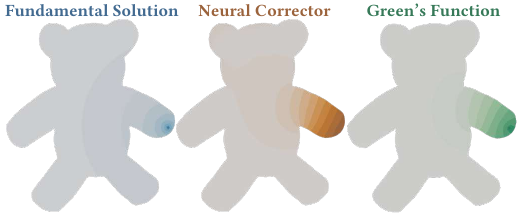}
    \captionsetup{skip=0pt}
  \caption{
  The analytic-corrector decomposition of the Neumann Laplace Green's function uses the analytic Green's function to capture the sharpness of the Green's function, letting our neural field capture an additional smooth corrector field required to enforce the 0-Neumann boundary conditions.
  }
  \Description{Decomposition of the Neumann Laplace Green's function into a fundamental solution and a neural corrector.}
  \label{fig:green-decomposition-bear}
\end{figure}
Since humanity's earliest days, our primal instinct when encountering an unknown object has remained unchanged: poke it and see what happens. 
It's no surprise then that a fundametal tool in the analysis of the Partial Differential Equations (PDEs) that model our world, is their response to a poke, or an impulse.
This response to an impulse is known as the \emph{Green's function} of the PDE operator, and in Computer Graphics they have proven to be useful for controlling and manipulating 3D geometry \cite{lipman2008greencoordinates,chen2023somiglianacoordinates}, shape analysis \cite{sun2009aconciseprovidablyinformativemultiscalebasedonheatdiffusion} and for accelerating PDE solvers \cite{chen2025lightning,james1999artdefo}.

While Green’s functions admit closed-form expressions for a small class of PDEs on simple, unbounded domains, computing them on complex geometries typically requires finite element discretization. This dependence on meshing becomes a major bottleneck as modern graphics pipelines increasingly rely on unstructured, high-resolution representations such as point clouds, neural fields, and constructive solid geometry, often with millions of primitives. For such data, meshing is costly or infeasible, making direct computation of Green’s functions intractable.

Despite recent progress on mesh-free PDE solvers using neural fields, relatively little work has addressed the problem of learning Green’s functions directly on unstructured function spaces. This is due to several challenges unique to Green’s functions. Suitable training objectives are unclear: existing approaches often rely on mesh-based supervision or residual minimization. Green's functions additionally are particularly sensitive to the imposed boundary conditions, which are particularly difficult to enforce in neural field formulations. In addition, Green’s functions exhibit sharp, localized structure that is difficult to represent with smooth function spaces. Finally, many applications require frequent evaluation of Green's functions across multiple impulse locations, and so our neural field needs to be capable of generalizing to arbitrary impulse locations.

We propose \emph{Variational Green’s Functions (VGF)}, an unsupervised method for learning Green’s functions of linear, elliptic, self-adjoint PDE operators directly on unstructured function spaces such as neural fields. To the best of our knowledge, we are the first to train a Green’s function neural field by minimizing the variational energy associated with the Green’s function PDE. This formulation requires no mesh-based supervision, depends only on first-order derivatives, and provides a natural mechanism for enforcing Neumann boundary conditions.

To capture the sharp structure of Green’s functions, we factor out the analytic free-space singularity and learn only a smooth corrective field, and we exploit the sifting property of the Dirac delta to accurately evaluate its contribution to the variational energy. Finally, to enable generalization across impulse locations, we introduce a hyperintegration-based training scheme that jointly samples spatial evaluation points and impulse locations at each epoch, leading to faster convergence and a single model capable of producing the results for arbitrary impulse locations, without re-training as shown in \reffig{batfem-comparison}.

For the Laplace, screened Poisson, and biharmonic operators, our method is capable of providing accurate Green's functions in real-time for arbitrary source locations..
We showcase further benefits of our neural field formulation by evaluating Green's functions on various geometric representations \reffig{teaser}, Green's functions that generalize over shape spaces, and inverse problems dependent on Green's functions, such as skinning weight design and shape segmentation.

%% file: related_work_shorter.tex
\section{Related Work}
Our method builds on two complementary lines of research: the use of Green's functions in geometry processing, and mesh-free approaches for solving partial differential equations using neural fields.
\subsection{Green's Functions in Computer Graphics}
Green's functions have long played a central role in accelerating PDE solvers and enabling localized, physically meaningful control in computer graphics. They form the foundation of the Boundary Element Method (BEM), which reduces volumetric PDEs to boundary integrals and has been applied to elastic deformation, surface-only fluid simulation, Poisson problems, and diffusion-based texture synthesis
\cite{james1999artdefo,sugimoto2022surfaceonlydynamicdeformablesusingaboundaryelementmethod,da2016surfaceonlyliquids,chen2025lightning,bang2023anadaptivefastmultipoleacceleratedhybridboundaryintegralequationmethodforaccuratediffusioncurves,sun2012diffusioncurvetexturesforresolutionindependantexturemapping}.
Despite their accuracy, BEM-based methods rely on clean, high-quality surface meshes, which are not always available in modern geometric pipelines.

Beyond solver acceleration, Green's functions have been used as controllable primitives for simulation and modeling. Examples include vortex-based smoke control via the Biot--Savart law \cite{angelidis2006acontrollablefastandstablebasisforvortexbasedsmokesimulation}, interactive elastic sculpting tools based on free-space elasticity kernels \cite{degoes2017regularizedkelvinlets,degoes2018dynamickelvinlets,chen2022gogreen}, and cage-based deformation using Laplace or elasticity Green's functions \cite{lipman2008greencoordinates,chen2023somiglianacoordinates}. In shape analysis, Green's functions of diffusion and wave operators underpin intrinsic descriptors and distance constructions
\cite{aubry2011wavekernel,sun2009aconciseprovidablyinformativemultiscalebasedonheatdiffusion,lipman2010biharmonicdistance}.

Our work builds on these insights while removing the reliance on explicit surface or volumetric discretizations. By learning Green's functions as neural fields, we extend localized, Green's-function-based reasoning to arbitrary unstructured geometric representations.
\subsection{Mesh-Free PDE Solvers}
Motivated by the growing prevalence of unstructured geometric data, numerous mesh-free PDE solvers have been proposed. Point-based methods such as Smoothed Particle Hydrodynamics, Reproducing Kernel Particle Methods, Point Set Surfaces, and implicit Moving Least Squares define function spaces directly over particles or point clouds
\cite{gingold1977smoothed,liu1995reproducing,alexa2001pointsetsurfaces,levin1998approximationpower}, with recent extensions enabling robust contact handling \cite{du2024robustandartefactfreedeformablecontactwithsmoothsurfacerepresentations}. While mesh-free, these approaches still solve PDEs through particle-based discretizations and iterative time stepping, coupling solutions to a specific sampling of the geometry.

Monte Carlo Geometry Processing offers an alternative by interpreting Green's functions probabilistically, enabling grid-free Poisson solves with minimal surface information \cite{sawhney2020mcgp}. However, each query still requires stochastic evaluation rather than amortized inference.

Neural fields provide a complementary paradigm by parameterizing continuous physical quantities as neural networks, offering effectively infinite resolution and the ability to operate on arbitrary domains
\cite{sitzmann2020siren,chang2025shapespace,modi2024simplicits,sellan2023neuralstochastic}. Variational neural solvers cast PDEs as energy minimization problems \cite{e2017deepritz}, though enforcing boundary conditions remains challenging in practice \cite{berrone2023enforcing}. Recent work addresses this by hard-constraining the network architecture to exactly satisfy boundary conditions \cite{dodik2025rbs}.

Our method follows this variational philosophy but targets the Green's function of the PDE operator itself. Unlike approaches that learn solutions via finite-element supervision \cite{yoo2025neural,melchers2024neuralgreensoperatorsforparametricpdes} or minimize PDE residuals on background grids \cite{teng2022learningGFNet}, our formulation learns Green's functions directly from the variational energy, without meshes, or grids and requires only distance-to-surface queries. This enables efficient, random-access evaluation and generalization across geometric parameterizations without re-discretization or re-solving.

%% file: method_otman.tex
\section{Green's Functions: A Variational Perspective}
\label{sec:green-variational}

Our framework relies on constructing a variational principle for the Green's functions of linear elliptic PDE operators.
The Green's function $G(x, s)$ of a PDE operator $\mathcal{L}$, is defined as the response of the operator at point $x \in \Rn{d}$ to an impulse source applied at $s \in \Rn{d}$ and satisfies the equations:
\begin{align}
\label{eq:generic-pde}
\mathcal{L} G(x, s) &= \delta(x, s) \quad  \forall x \in \Omega, \nonumber \\
    G(x, s) &= h(x) \quad  \forall x \in \partial \Omega_d,\\
    \partial_n G(x, s) &= g(x) \quad  \forall x \in \partial \Omega_n, \nonumber
\end{align}
where $\mathcal{L}$ is a linear self-adjoint elliptic PDE operator of order $2d \,\forall d \in \mathbb{N}$, $\delta(x, s)$ is a Dirac delta function centered at $s$, $h(x)$ are Dirichlet boundary conditions, and $g(x)$ are Neumann boundary conditions. 
Furthermore, if the PDE operator is of order greater than $2$, we assume an additional homogeneous boundary condition on the $d$-th derivative of the Green's function, to make the problem well-posed.

\begin{figure}[t]
    \centering
    \captionsetup{skip=0pt}
    \includegraphics[width=\linewidth]{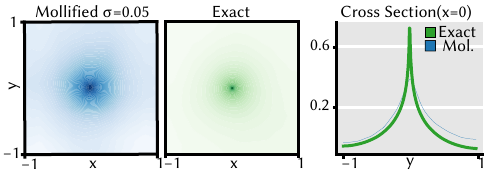}
    \caption{Leveraging the sifting property of the Dirac delta function is essential for capturing the sharpness of the impulse source.}
    \label{fig:sifting-property}
\end{figure}

Such linear elliptic PDE operators—including the Laplace, screened Poisson, and biharmonic operators—admit a variational formulation 
\begin{align}
\label{eq:generic-variational}
G(x, s) = &\argmin_{u(x)}  E_{\mathcal{L}}[u(x)] - \left\langle \delta(x, s), u(x) \right\rangle \nonumber \\
& \text{s.t.} \quad u(x) = h(x) \quad \forall x \in \partial \Omega_d,  \\
& \text{and} \quad \partial_n u(x) = g(x) \quad \forall x \in \partial \Omega_n. \nonumber
\end{align}
provided the function space for $G(x, s)$ is sufficiently smooth, as guaranteed by the Riesz Representation Theorem \cite{riesz1907systemes}.
Above, $E_{\mathcal{L}}[u(x)]$ is the variational energy functional associated to the PDE operator $\mathcal{L}$, generically given by:
\begin{align}
\label{eq:generic-energy}
E_{\mathcal{L}}[u(x)] = \frac{1}{2} \left\langle\mathcal{L} u(x), u(x)\right\rangle . 
\end{align}
To perfectly capture the sharp dirac delta source term in our variational formulation, we leverage the sifting property of the dirac delta function,$
\left\langle \delta(x, s), u(x) \right\rangle =  u(s)$.
This provides us with an expression for evaluating contribution of the source term to the energy functional requiring only a simple evaluation of $u(s)$. 
Leveraging this sifting property is essential for ensuring that our final energy landscape reflects the sharpness of the impulse source, as shown in \reffig{sifting-property}. 

The advantage of casting the Green's functions as the solution to the variational problem \refeq{generic-variational}, as we will see, is that it will allow us to define an unsupervised objective function which we can use to optimize the parameters of arbitrary function representations, including neural fields. 
We now turn our attention to the specific variational formulations for the Green's functions of the Laplace, screened Poisson, and biharmonic operators.

\subsection{Green's Functions of the Laplace Operator $\Delta$}
\label{sec:green-laplace}
The first PDE operator we consider is the Laplace operator $\mathcal{L} = \Delta$.
We consider a domain $\Omega$ as an open subset of $\mathbb{R}^n$. 
Within this domain, the Laplacian operator $\Delta$ is defined intrinsically as the divergence of the gradient, $\nabla \cdot \nabla$.

For the Laplace operator, the corresponding variational energy functional can be rewritten via integration by parts into the Dirichlet energy\cite{wardetzky2011discrete}:
\begin{align}
    E_{\Delta}[u(x)] = \frac{1}{2}  || \nabla u(x)|| ^2 - \int_{\partial \Omega_n} u(x) g(x) dx,
\end{align}
where the second term is the contribution to the functional arising from the Neumann boundary conditions, if they exist.
This second term allows us to enforce the Neumann boundary conditions \emph{naturally}  within our energy functional \cite{stein2018natural}, without requiring any additional constraint on the function space.
In particular, for vanishing Neumann boundary conditions, that second term disappears, and the boundary conditions arise naturally from minimizing the first term alone.

\begin{figure}[t]
    \centering
\includegraphics[width=1\columnwidth,keepaspectratio]{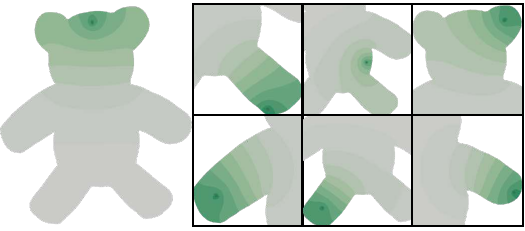}
    \captionsetup{skip=0pt}
    \caption{ 
        Minimizing the Dirichlet energy naturally enforces vanishing Neumann boundary conditions in our Green's functions for the Laplace operator. These boundary conditions are preserved across various impulse sources all generated by the \emph{same} neural field.
    }
    \label{fig:boundary-condition-poisson}
  \end{figure}

\paragraph{Modifying the Source for Neumann Boundary Conditions}
For the Laplace Green’s function $u(x)=G(x,s)$ with Neumann boundary conditions, the standard variational formulation does not admit a solution for arbitrary prescribed flux $g(x)$. By the divergence theorem, any solution of
$\Delta u(x)=f(x)$ must satisfy
\begin{equation}
    \int_{\partial \Omega_n} \partial_n u(x)\,dx = \int_{\Omega} f(x)\,dx .
\end{equation}
When $f(x)=\delta(x,s)$, the right-hand side evaluates to $1$, while the left-hand side is fixed by the prescribed Neumann data $\partial_n u(x)=g(x)$. This compatibility condition generally fails for arbitrary $g(x)$.

To make the problem well posed, we follow the lead of \citet{choksi2022partial} and modify the source with a background sink term,
$$
    f(x) = \delta(x,s) + \frac{1}{|\Omega|}
    \left( \int_{\partial \Omega_n} g(x)\,dx - 1 \right),
$$
which enforces the required flux balance. This modification adds a constant term to the energy functional,
\begin{equation}
    E_s[u] =
    \frac{1}{|\Omega|}
    \left( \int_{\partial \Omega_n} g(x)\,dx - 1 \right).
\end{equation}

\paragraph{Removing the Nullspace Ambiguity}
When no Dirichlet boundary conditions are present, the Laplace operator admits a constant nullspace, making the solution unique only up to an additive constant. We remove this ambiguity by enforcing a zero-mean constraint, implemented either by projection or by adding the penalty
\begin{equation}
    E_u[u] =
    \left(
        \frac{1}{|\Omega|}
        \int_{\Omega} u(x)\,dx
    \right)^2 .
\end{equation}

\subsection{Green's Functions of the Screened Poisson Operator $k^2 - \Delta $}
Now that we have a variational formulation for the Green's functions of the Laplace operator with Dirichlet or Neumann boundary conditions, we can easily extend this framework to the screened Poisson operator $\mathcal{L} = k^2 - \Delta$, where $k > 0$ is a screening parameter.
This operator has a well known variational energy functional \cite{kazhdan2013screened}, resembling the Dirichlet energy:
\begin{align}
    E_{k^2 - \Delta}[u(x)] = \frac{k^2}{2}  ||u(x)||^2 +  \frac{1}{2} || \nabla u(x) ||^2 - \int_{\partial \Omega_n} u(x) g(x) dx.
\end{align}
The third term is the contribution to the functional arising from the Neumann boundary conditions, as was the case for the Dirichlet energy.
Unlike the Laplace operator, the screened Poisson operator has no nullspace ambiguity, and so we can directly minimize the energy functional \refeq{generic-variational} to obtain our Green's function.

\begin{figure}[t]
  \centering
  \includegraphics[width=1\columnwidth,keepaspectratio]{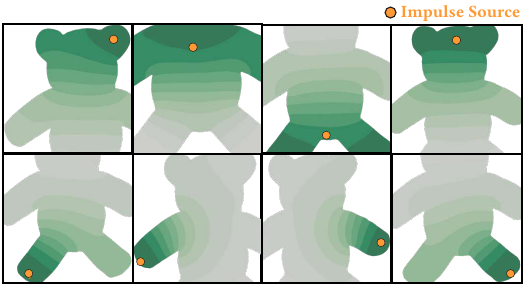}
      \captionsetup{skip=0pt}
  \caption{
  Vanishing Neumann boundary conditions can be enforced naturally by our two-stage mixed method to solving for the Green's functions of the biharmonic operator. These boundary conditions are preserved across various queried impulse source locations, all generated by the same pre-trained neural field. 
  }
  \label{fig:boundary-condition-biharmonic}
\end{figure}

\subsection{Green's Functions of the Biharmonic Operator $\Delta^2$}

The biharmonic operator admits a classical variational energy, often referred to as the Laplace energy functional \cite{stein2018natural}:
\begin{align}
E_{\Delta^2}[u] = \frac{1}{2}\|\Delta u\|^2
- \int_{\partial \Omega_\Delta} p(x)\,\partial_n u(x)\,dx
+ \int_{\partial \Omega_{\partial \Delta}} q(x)\,u(x)\,dx . \nonumber
\end{align}
However, the boundary terms in this formulation prescribe values of $\Delta u$ and $\partial_n \Delta u$ on the boundary, rather than directly controlling the Neumann boundary conditions of $u$ itself. As a result, this energy cannot be used directly to enforce the desired boundary conditions for biharmonic Green’s functions.

\paragraph{Mixed Formulation}
To address this limitation, we adopt a mixed formulation of the biharmonic operator \cite{jacobson2010mixed}. The key observation is that the bilaplacian factors into two Laplacians,
\begin{equation}
\Delta^2 = (-\Delta)(-\Delta),
\end{equation}
allowing the fourth-order problem to be decomposed into two coupled second-order problems.

This leads to a simple two-stage construction of the biharmonic Green’s function. First, we solve a Laplace Green’s function problem with zero Neumann boundary conditions:
\begin{equation}
\label{eq:biharmonic-step1}
\textbf{Step 1:}\quad
\begin{cases}
-\Delta v(x,s) = \delta(x,s) - \frac{1}{|\Omega|}, & x \in \Omega, \\
\partial_n v(x,s) = 0, & x \in \partial \Omega .
\end{cases}
\end{equation}
This stage ensures that the Laplacian of the final Green’s function satisfies a compatible boundary condition, which is required for solvability of the biharmonic problem.

In the second stage, we recover the biharmonic Green’s function by solving a Poisson problem with $v(x,s)$ as the source:
\begin{equation}
\label{eq:biharmonic-step2}
\textbf{Step 2:}\quad
\begin{cases}
-\Delta G(x,s) = v(x,s), & x \in \Omega, \\
G(x,s) = h(x), & x \in \partial \Omega_d, \\
\partial_n G(x,s) = g(x), & x \in \partial \Omega_n .
\end{cases}
\end{equation}
This final stage enforces the desired Dirichlet or Neumann boundary conditions on the biharmonic Green’s function.

Both stages admit standard variational formulations. The first stage corresponds to a Laplace Green’s function with zero Neumann boundary conditions and a zero-mean constraint, while the second stage is a Poisson problem with arbitrary boundary conditions. In practice, we implement this mixed formulation using two neural fields: the first predicts $v(x,s)$, and the second predicts $G(x,s)$ conditioned on the output of the first.

\paragraph{Choice of Intermediate Boundary Conditions}
The mixed formulation is not unique, as different boundary conditions may be imposed on the auxiliary field $v(x,s)$. For example, one could alternatively enforce Dirichlet conditions on $v$ along $\partial\Omega$, leading to a different but still valid biharmonic formulation.

We choose zero Neumann boundary conditions for $v(x,s)$ for two practical reasons. First, this choice is simple to implement within our variational framework: zero Neumann conditions arise naturally by omitting boundary integrals from the Dirichlet energy, as discussed in \refsec{green-laplace}. Second, it avoids introducing additional boundary constraints or penalty terms, allowing the intermediate Green’s function to be learned without explicit boundary supervision.

\section{Implicit Function Parameterization of Green's Functions}
Now that we have a variational formulation for the Green's functions of the Laplace, screened Poisson, and biharmonic operators, we need to define a functional representation for our Green's functions.
We start with a generic functional parameterization:
\begin{align}
\label{eq:generic-green}
G(x, s) = G_{\theta}(x, s)
\end{align}

\begin{figure}[t]
  \centering
  \includegraphics[width=1\columnwidth,keepaspectratio]{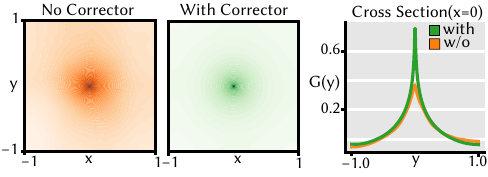}
  \captionsetup{skip=0pt}
  \caption{
    The analytic corrector decomposition allows our final learned Green's function to properly capture the sharp singularity characteristic of the Laplacian's Green's function on a rectangular domain.
  }
  \Description{
      Corrector Comparison
  }
  \label{fig:green-decomposition}
\end{figure}

Where $\theta$ are the parameters of our function to be fitted via a minimization of the variational energy functional (e.g. the weights of a neural network, the coefficients of a polynomial, etc.), specific to the PDE operator we are considering.
We require $G_{\theta}(x, s)$ to be: (1) Once differentiable with respect to $\theta$ to allow for gradient-based optimization (2) Capable of capturing sharp singularities characteristic of Laplacian Green functions, and (3) amenable to enforcing Dirichlet boundary conditions. 
The first criterion (1) is trivially enforced via our choice of a neural field parameterization. We move on to satisfying the remaining criteria.

\subsection{Fundamental Solution Decomposition}
To ensure our function space can correctly resolve the sharp singularity of the solution near the source point, we further decompose our function space into an analytic fundamental solution component and a fitted "corrector" component:
\begin{align}
\label{eq:generic-green-decomposition}
G_{\theta}(x, s) = \Phi(x, s) + H_\theta(x, s) 
\end{align}

The analytic fundamental solution component, $\Phi(x, s)$, corresponds to the free-space solution of the PDE operator, and is essential for capturing the sharp singularity perfectly.
In contrast, the corrector component, $H_\theta(x, s)$, is a function that is fitted to make the overall solution aware of the specific geometry defined by the boundary of the domain.
This decomposition ultimately allows our framework to capture sharp singularities characteristic of Green's functions, while being sensitive to the specific geometry of the domain as shown in \reffig{green-decomposition} and \reffig{green-decomposition-bear}.

Analytic fundamental solutions for the three PDE operators considered are given in \reftab{fundamental-solutions}.
\begin{table}[t]
    \centering
        \captionsetup{skip=0pt}
    \caption{Fundamental solutions for the three PDE operators considered.}
    \label{tab:fundamental-solutions}
    \renewcommand{\arraystretch}{1.3}
    \begin{tabular}{|c|c|c|}
    \hline
    \textbf{PDE Operator} & $\Phi(x,s),  \textbf{2D}$ & $\Phi(x,s),  \textbf{3D}$ \\
    \hline
    \rowcolor{green!10}
     ($\Delta$) & $-\dfrac{1}{2\pi} \ln \|x - s\|$ & $\dfrac{1}{4\pi \|x - s\|}$ \\[0.5em]
    \hline
    \rowcolor{white}
    ($k^2 - \Delta$) & $\dfrac{1}{2\pi} K_0\left(k\|x - s\|\right)$ & $\dfrac{\exp(-k\|x - s\|)}{4\pi \|x - s\|}$ \\[0.5em]
    \hline
    \rowcolor{green!10}
     ($\Delta^2$) & $\dfrac{1}{8\pi} \|x - s\|^2 \ln\|x - s\|$ & $\dfrac{\|x - s\|}{8\pi}$ \\[0.5em]
    \hline
    \end{tabular}
\end{table}

\subsection{Enforcing Dirichlet Boundary Conditions}
\label{sec:green-dirichlet}
Standard penalty-based methods for enforcing boundary conditions often struggle with balancing interior smoothness against boundary adherence \cite{berrone2023enforcing}.
Instead, we employ a hard enforcement strategy via reparameterization, following the method of \citet{dodik2025rbs}, constructing a field $\tilde{G}_\theta$ that satisfies the boundary conditions by construction.
We define the reparameterized Green's function as:
\begin{equation}
\label{eq:reparam}
\tilde{G}_\theta(x,s) = w(x) h(x) + (1 - w(x)) G_\theta(x,s),
\end{equation}
where $w(x)= w(d(x))$ is a function of the distance $d(x)$  from $x$ to the nearest boundary, and $w(d)$ is a smooth blending function satisfying $w(0)=1$ and $w(d)=0$ for $d > \epsilon$.
Near the boundary (where $w \approx 1$), the network output is replaced by the required boundary data $h(x) $.
In the interior (where $w = 0$), the network has full control to minimize the energy.

To ensure smoothness for gradient computation, we use the blending curve:
\begin{equation}
w(d(x)) = \left( 1 - \frac{d(x)^2}{\epsilon^2}\right)^2,
\end{equation}
whose squared terms guarantee vanishing derivatives at both $d=0$ and $d=\epsilon$.
This reparameterization ensures the Green's function satisfies the boundary conditions by construction, and allows us to train the Green's function in an unsupervised manner.

\section{Hyperintegration Sampling}
\label{sec:hyperintegration-sampling}

Now that we have a variational formulation for the Green's functions, as well as a decoupled parameterization capable of capturing sharp features, we can discuss how to evaluate the energy functional to fit our function space parameters $\theta$.

Using the bilinear energy density $\psi_{\mathcal{L}}(u, v) = \frac{1}{2} u \mathcal{L}u -  \frac{1}{|\Omega|}v,$
we can rewrite \refeq{generic-variational} and define our Green's function as the minimizer of the energy functional:
\begin{align}
    G(x, s) = \argmin_{u(x)}  \int_{\Omega} \psi_{\mathcal{L}}(u(x), u(s)) dx.
\end{align}
From this form, to fit our parameterized Green's function $u(x) = G_\theta(x, s)$ to a single fixed source location $s=s_0$, we minimize the variational energy to solve for the parameters $\theta$:
\begin{align}
    \theta = \argmin_{\theta}  \int_{\Omega} \psi_{\mathcal{L}}(G_\theta(x, s_0), G_\theta(s_0, s_0)) dx.
\end{align}
Departing from the single impulse case, we want our parameters $\theta$ to provide accurate Green's functions for source impulses applied at \emph{any} location in our domain $\Omega$. 
A naive strategy would be to sample a single source location, take an optimizer step to fit the neural field assuming this fiixed source location, and repeating this process by changing the source location every epoch. 
We call this strategy \emph{single-source integration} and we've found that it leads to slower and less stable convergence.

\begin{figure}[t]
  \centering
      \captionsetup{skip=0pt}
  \includegraphics[width=1\columnwidth,keepaspectratio]{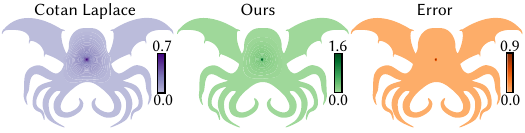}
  \caption{Our VGF's for the Laplace equation can be made to enforce Dirichlet boundary conditions  via an output reparameterization of our neural field. The main source of error is sharply localized within the singularity. }
  \label{fig:poisson-dirichlet}
\end{figure}
\begin{figure}[b]
    \includegraphics[width=1\columnwidth,keepaspectratio]{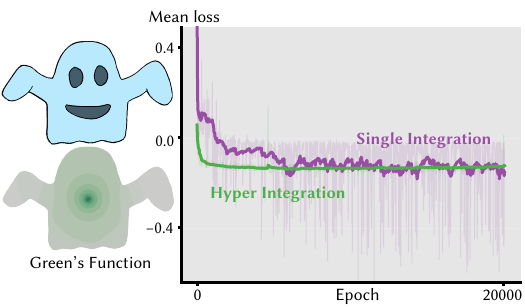}
        \captionsetup{skip=0pt}
    \caption{Hyper integration allows our neural field training parameters to quickly converge, reaching a stable plateau, whereas single-source integration requires many more epochs to converge.}
    \Description{Hyper Integration vs Single Integration}
    \label{fig:hyperintegration}
  \end{figure}

Instead, we develop \emph{hyperintegration sampling}, which jointly samples $N$ random evaluation points and source locations every epoch.  
 We integrate the objective over all possible source locations $s \in \Omega$, to that our $\theta$ generalizes to arbitrary source locations:
\begin{align}
    \theta = \argmin_{\theta} \iint_{\Omega \times \Omega} \psi_{\mathcal{L}}(G_\theta(x, s), G_\theta(s, s)) dx ds.
\end{align}
We estimate this double integral every epoch via stochastic cubature, drawing $N$ samples $\{x_i, s_i\}_{i=1}^{N}$ randomly within the product space $\Omega \times \Omega$, providing us with the estimated loss every epoch:
\begin{align}
    L(\theta) = \frac{1}{N} \sum_{i=1}^N \psi_{\mathcal{L}}(G_\theta(x_i, s_i), G_\theta(s_i, s_i)).
\end{align}
Where every epoch we sample a new set of $N$ random samples $\{x_i, s_i\}_{i=1}^{N}$ and update our parameters $\theta$ incrementally.
We show in \reffig{hyperintegration} that hyperintegration sampling leads to faster and more stable convergence than single-source integration.

%% file: experiments.tex
\section{Experiments}
\label{sec:experiments}

We implement our method in PyTorch. Unless otherwise specified, all experiments were conducted on a Macbook Pro with an Apple M4 Max chip and 128GB of RAM.

We use a 4-layer, 256-unit multilayer perceptron (MLP) with sine activation functions and positional encoding ($2^6$ frequencies). To train our neural field, we use the Adam optimizer with a learning rate of $10^{-3}$ for 50k epochs, with early stopping triggered after a patience of 2k epochs.

\subsection{Training time \& Statistics}
We collect training times, inference times, and error statistics for various examples in our paper and summarize them in Table \ref{tab:training-stats}. Observe that evaluating our Green's functions over inference can be done at interactive rates, taking a maximum of $4.9\text{ms}$, which we use to evaluate Green's functions over various impulses \reffig{batfem-comparison}, shapes in a shape space \reffig{shape-space}, or PDE parameters \reffig{screened-poisson-neumann-2}.

Furthermore, observe that our method beats the baseline provided by \citet{sharp2020robustlaplacian} for all 3D examples, whereas their method provides an almost perfect result on 2D planar domains.
\begin{table*}[t]
    \centering
        \captionsetup{skip=0pt}
    \caption{Training (train) and inference (inf.) times for the examples presented in this paper, along with the number of vertices per shape and the MLP parameters ($\#$params). The inference time is the average duration required to evaluate a source point against all vertices, computed over a 1000 uniformly sampled sources. The mean error baseline we compare to is computed from the robust Laplacian \cite{sharp2020robustlaplacian}.}
    \label{tab:training-stats}
    \renewcommand{\arraystretch}{1.3}
    \begin{tabular}{|
        >{\centering\arraybackslash}p{0.16\linewidth}|
        *{6}{>{\centering\arraybackslash}p{0.08\linewidth}|}
    }
    \hline
    \textbf{example} & \textbf{\# vertices} &\textbf{train} & \textbf{inf.}  & \textbf{\# params} & \textbf{mean err.} & \textbf{mean err.} \\
    & & \textbf{(min)} & \textbf{(ms)}  &  & \textbf{base(\%)}& \textbf{ours (\%)} \\
    \hline
    \rowcolor{green!10}
    Cthulhu & $4.5\times10^{3}$ & 22 & $4.9\times10^{1}$ & $2.00\times10^{4}$ & \textbf{0.0071} & 0.58 \\
    \hline
    \rowcolor{white}
    Crab & $5.7\times10^{4}$ & 51 & $3.9\times10^{1}$ & $2.84\times10^{5}$ & 3.56 & \textbf{1.61} \\
    \hline
    \rowcolor{green!10}
    Sappho & $1.1\times10^{4}$ & $1.5\times10^{2}$ & $1.8\times10^{1}$ & $5.67\times10^{5}$ & 27.38 & \textbf{4.80} \\
    \hline
    \rowcolor{white}
    Koala & $8.1\times10^{3}$ & $1.2\times10^{2}$ & $1.7\times10^{1}$ & $5.67\times10^{5}$ & 19.47 & \textbf{3.85} \\
    \hline
    \rowcolor{green!10}
    Octopus & $4.8\times10^{3}$ & $1.5\times10^{2}$ & $1.5\times10^{1}$ & $1.10\times10^{6}$ & 9.27 & \textbf{6.52} \\
    \hline
    \rowcolor{white}
    Teddy Bear (3D) & $4.9\times10^{4}$ & 84 & $3.6\times10^{1}$ & $2.84\times10^{5}$ & 1.32 & \textbf{0.27} \\
    \hline
    \rowcolor{green!10}
    Teddy Bear (2D) & $4.6\times10^{3}$ & 78 & $9.0\times10^{0}$ & $2.77\times10^{5}$ & \textbf{0.0020} & 0.27 \\
    \hline
    \rowcolor{white}
    Bunny & $7.8\times10^{3}$ & $2.8\times10^{2}$ & $1.9\times10^{1}$ & $4.84\times10^{5}$ & 8.02 & \textbf{7.95} \\
    \hline
    \rowcolor{green!10}
    Frog & $1.1\times10^{4}$ & $4.6\times10^{2}$ & $2.0\times10^{1}$ & $5.67\times10^{5}$ & 15.43 & \textbf{8.12} \\
    \hline
    \rowcolor{white}
    Bat & $9.3\times10^{3}$ & 54 & $1.5\times10^{1}$ & $2.84\times10^{5}$ & 7.66 & \textbf{1.27} \\
    \hline
    \end{tabular}
\end{table*}

\section{Results and Applications}
\label{sec:results-and-applications}

\begin{figure}[t]
    \centering
  \includegraphics[width=1\columnwidth,keepaspectratio]{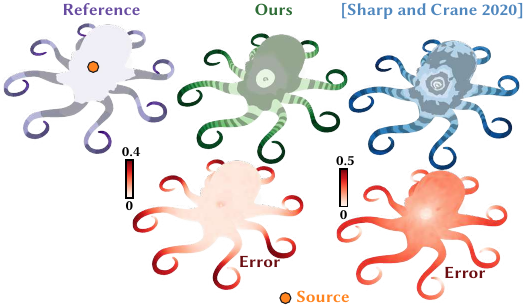}
      \captionsetup{skip=0pt}
    \caption{Commute time distance from one source point on an octopus tentacle, to all other points on the octopus. Our method doesn't rely on a spectral approximation to compute commute time distance, providing a result closer to the ground truth obtained via the Green's functions of the discrete cotan Laplacian.}
    \Description{}
    \label{fig:commute-time}
  \end{figure}

\subsection{Spectral Distances}
\label{sec:spectral-distances}
Spectral distances measure intrinsic surface similarity and are robust to noise. They are typically defined via Laplacian or biharmonic Green's functions, but computing these exactly is expensive and often approximated via spectral decomposition \cite{lipman2010biharmonicdistance}. Our method computes these distances efficiently without such approximations.
\paragraph{Commute Time Distance}
The commute time distance $d_{CT}(x, y)$ measures the expected round-trip time of a random walker:
\begin{equation}
d_{CT}^2(x, y) = G_\Delta(x, x) + G_\Delta(y, y) - 2G_\Delta(x, y).
\end{equation}
Traditional computation is costly and non-differentiable. With VGF, it reduces to a simple network query, reproducing higher-fidelity distances than spectral approximations even with $1000$ eigenvectors \cite{sharp2020robustlaplacian}.

\paragraph{Biharmonic Distance}
The biharmonic distance $d_{Bi}(x, y)$ uses the biharmonic Green's function:
\begin{equation}
d_{Bi}^2(x, y) = G_{\Delta^2}(x, x) + G_{\Delta^2}(y, y) - 2G_{\Delta^2}(x, y).
\end{equation}
It penalizes high-frequency variations, yielding smoother distances for shape matching. VGF reproduces these distances with lower error than the robust Laplacian \cite{sharp2020robustlaplacian}.

\subsection{Shape Segmentation via Biharmonic Clustering}
\label{sec:biharmonic-clustering}
We further exploit the differentiability of our biharmonic Green’s functions with respect to the impulse location to perform shape-aware clustering.
Figure~\ref{fig:biharmonic-clustering} illustrates the resulting segmentation.

Given $K$ cluster centers $\{c_k\}_{k=1}^K$ on the shape, each point $x$ is assigned to its nearest center under a biharmonic distance proxy, forming clusters $\{C_k\}$. We optimize the center positions by minimizing
\begin{equation}
    \mathcal{L} =
    \sum_{k=1}^K
    \left[
        G_{\Delta^2}(c_k,c_k)
        - \frac{1}{|C_k|}
        \sum_{x \in C_k}
        \bigl(G_{\Delta^2}(x,c_k) + G_{\Delta^2}(c_k,x)\bigr)
    \right].
\end{equation}
The first term encourages centers to move toward geometrically salient regions, while the second promotes compact clusters under the biharmonic metric.

Optimization alternates between two steps: (i) assigning each point to its nearest center using the distance proxy
$d^2(x,c_k) \approx G_{\Delta^2}(c_k,c_k) - (G_{\Delta^2}(x,c_k) + G_{\Delta^2}(c_k,x))$, and
(ii) updating center locations via Adam, with gradients obtained through automatic differentiation.

As shown in \reffig{biharmonic-clustering}, centers initialized in a clustered configuration quickly spread to produce a balanced segmentation. This continuous optimization over impulse locations is difficult to achieve with traditional mesh-based methods, where impulses are restricted to discrete vertices and gradients are ill-defined.

\subsection{Biharmonic Deformation Weights}
\label{sec:biharmonic-deformation}
The biharmonic equation is widely used to construct smooth skinning weights for deformation and character rigging \cite{wang2015linearsubspacedesign,dodik2025rbs,boundedbiharmonicweights2011}. We show that biharmonic Green’s functions provide a natural tool for the inverse design of such weights, enabling the fitting of skinning weights that best reproduce a target deformation.

Starting from the zero-Neumann biharmonic Green’s function, we construct a normalized skinning weight
\begin{equation}
    W(x,s) =
    \frac{G_{\Delta^2}(x,s) - \min_y G_{\Delta^2}(y,s)}
         {\max_y G_{\Delta^2}(y,s) - \min_y G_{\Delta^2}(y,s)} .
\end{equation}

The weight $W(x,s)$ is attached to a skinning handle represented by an affine transform $\mat{T} \in \mathbb{R}^{3 \times 4}$, producing a displaced position
\[
    u(x) = W(x,s)\,\mat{T}[x,1]^T .
\]
\reffig{biharmonic-deformation} shows a localized deformation on a high-resolution crab mesh produced using a biharmonic Green’s function generated by our method. Because Green’s functions can be evaluated for arbitrary impulse locations, our approach enables rapid prototyping of localized skinning weights for interactive rigging and posing.

\begin{figure}
    \centering
        \captionsetup{skip=0pt}
    \includegraphics[width=\linewidth]{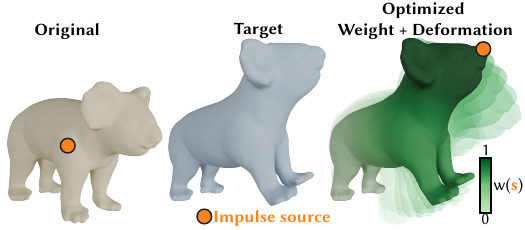}
    \caption{We jointly optimized for an impulse position, which parametrizes a biharmoinc skinning weight, as well as a skinning transformation applied to that skinning weight, in order to match a target deformation.}
    \label{fig:inverse-deformation}
\end{figure}

The differentiability of our Green’s functions enables direct solution of inverse deformation problems. Given an observed deformation $u_{\text{obs}}(x)$, we jointly optimize for the impulse location $s$ and the skinning transformation $\mat{T}$ that best reproduce the deformation 
$ \argmin_{s, \mat{T}} \sum_{x \in \Omega}
    \left\| u_{\text{obs}}(x) - W(x,s)\,\mat{T}[x,1]^T \right\|^2 $.

We solve this objective using gradient descent with ADAM \cite{kingma2014adam}. Since the skinning weight $W(x,s)$ is differentiable with respect to the impulse location $s$, gradients can be computed directly via automatic differentiation. This stands in contrast to traditional discrete methods, where impulse locations are restricted to mesh vertices and are difficult to optimize continuously. As shown in \reffig{inverse-deformation}, starting from a random initialization, the optimizer successfully recovers both the impulse location and the associated deformation.
\subsection{Shape Space Green's Functions}
By virtue of adopting a neural field as a function space, we can easily condition our Green's functions on external parameters, such as various shape parameters \reffig{shape-space}, or various screening parameters \ref{fig:screened-poisson-neumann-2}.
To generalize over shape code, every epoch we sample a new $d$-dimensional shape parameter $\mathbf{z}$ and update our parameters $\theta$ incrementally to minimize the loss, using Adam.

This allows a single pre-trained neural field to quickly evaluate Green's functions upon inference while changing shape parameters of the PDE in real time.
This would normally incur an expensive recomputation step with standard mesh-based techniques.

%% file: conclusion.tex
\section{Limitations}

Our method inherits several limitations from neural fields as a function representation. First, gradients of the solution can be noisy, particularly in regions where the Green's function exhibits rapid spatial variation. Second, geometries with pinches(positions that are close in euclidean space, but far geodesically) are challenging to resolve with our neural field function space. Furthermore, our natural boundary conditions are not guaranteed to be satisfied if the neural field hasn't converged.  Finally, our variational formulation currently only describes Dirichlet and Neumann boundary conditions; Robin boundary conditions are not yet supported.

\section{Conclusion}
By leveraging variational principles for PDE operators, we can train Green's function neural fields in an unsupervised manner that resolve sharp features, obey Neumann boundary conditions, and that generalize across impulse locations and parameters conditioning the PDE operator.

We are excited about several promising directions for future work: Extending our method to curved surfaces with intrinsic curvature would enable applications in differential geometry and manifold learning. 
Applying our variational framework to elasticity and dynamics presents an exciting opportunity to model time-dependent phenomena and material behavior. Investigating different function spaces may yield more efficient or expressive representations. Perhaps most ambitiously, generalizing our approach to unseen shapes would enable truly universal Green's function models that transfer across diverse geometries.

%% file: figures.tex
\begin{figure}
  \includegraphics[width=1\columnwidth,keepaspectratio]{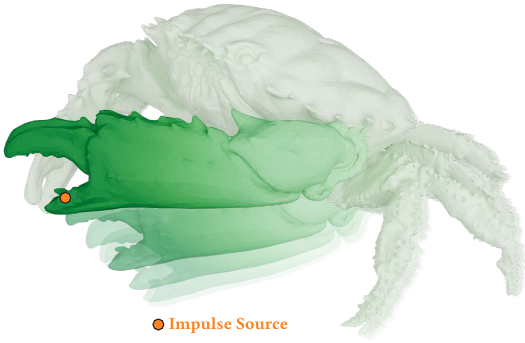}
  \caption{Our VGF for the biharmonic equation can be used to construct skinning weights, which we can use wtih Linear Blend Skinning to pose and articulate our crab. }
  \Description{Biharmonic Weights Deformation}
  \label{fig:biharmonic-deformation}
\end{figure}

\begin{figure}[t]
  \includegraphics[width=1\columnwidth,keepaspectratio]{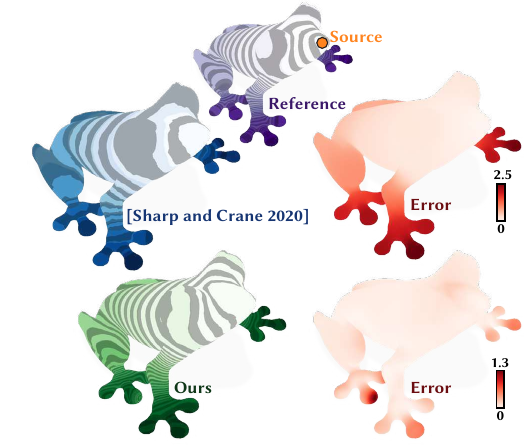}
  \caption{We compute biharmonic distances between points on a surface using our VGF. The distances are computed using the biharmonic Green's function, and achieve lower error when compared to distances computed using the robust Laplacian of \cite{sharp2020robustlaplacian}.}
  \Description{Biharmonic Distance}
  \label{fig:biharmonic-distance}
\end{figure}

\begin{figure}[t]
  \centering
  \includegraphics[width=1\columnwidth,]{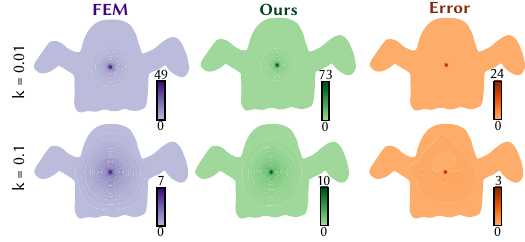}
  \caption{Our VGF's for the screened Poisson equation can be made to enforce Dirichlet boundary conditions  via an output reparameterization of our neural field. The main source of error is sharply localized within the singularity.}
  \Description{.}
  \label{fig:screened-dirichlet}
\end{figure}

\begin{figure}[b]
  \centering
  \includegraphics[width=1\columnwidth,keepaspectratio]{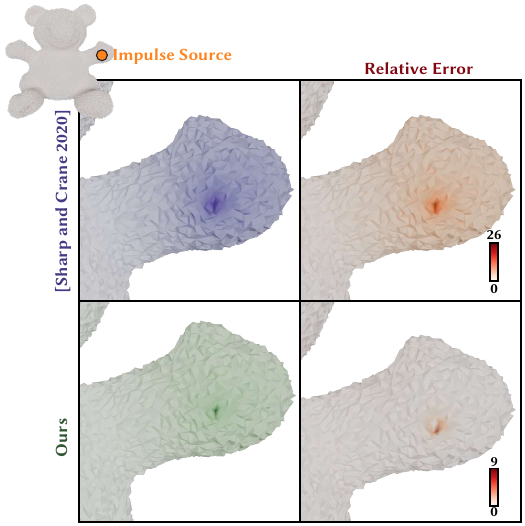}
  \caption{We solve the Green's function of the volumetric Poisson equation with vanishing Neumann boundary conditions on a teddy bear toy.
  Our VGF solution matches closely to the ground truth obtained from the cotan Laplace operator. In contrast, computing the Green's function via the robust Laplacian \cite{sharp2020robustlaplacian} incurs more error, further spread out over the entire domain.
  }
  \label{fig:poisson-neumann-error}
\end{figure}

\begin{figure}[]
  \centering
\includegraphics[width=1\columnwidth,keepaspectratio]{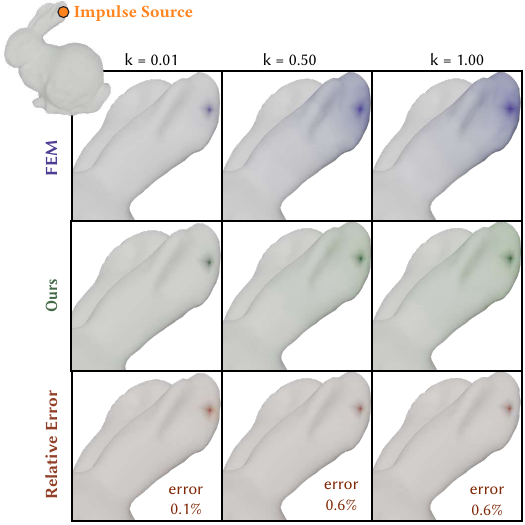}
  \caption{A single VGF neural field can produce solutions to the screened Poisson equation for various impulses \emph{and} various screening parameters. Our results match closely to the standard solution from the discrete screened Poisson operator obtained from the finite element method.}
  \Description{}
  \label{fig:screened-poisson-neumann-2}
\end{figure}

\begin{figure}[t]
  \centering
  \includegraphics[width=1\columnwidth]{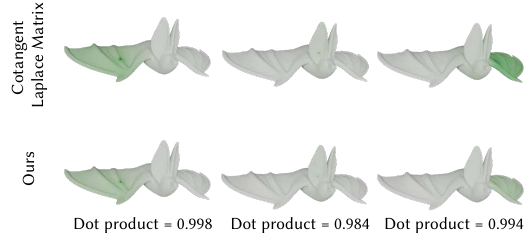}
  \caption{Our neural field reproduces Green's functions that closely match those of the discrete cotangent Laplacian in real time. A single trained model produces all results without retraining.}
  \label{fig:batfem-comparison}
\end{figure}

\begin{figure}[t]
  \centering
  \includegraphics[width=1\columnwidth,keepaspectratio]{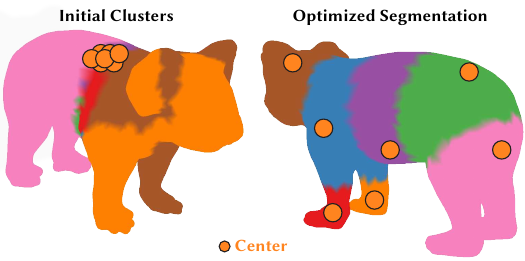}
  \caption{Shape Segmentation. We optimize for an initial set of clusters that are centralized. Over a few optimization steps, our method creates a balanced segmentation using our biharmonic clustering.}
  \Description{Shape Segmentation}
  \label{fig:biharmonic-clustering}
\end{figure}

\begin{figure}[b]
  \includegraphics[width=1\columnwidth,keepaspectratio]{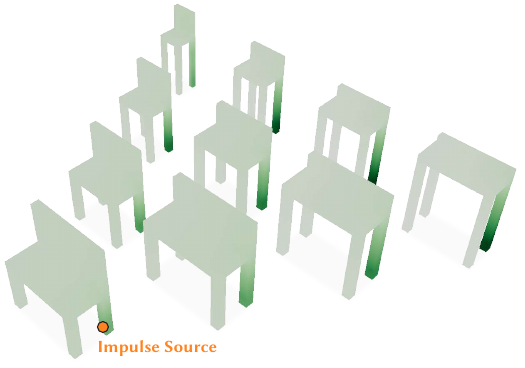}
  \caption{  \label{fig:shape-space}Green's Functions of the Laplace operator with zero-Neumann boundary conditions evaluated over a shape space \emph{without} re-training the neural field.}
  \Description{Green's Functions of the Laplace operator with zero-Neumann boundary conditions in a shape space}
  \label{fig:shape-space}
\end{figure}

\begin{figure}[t]
  \centering
\includegraphics[width=1\columnwidth,keepaspectratio]{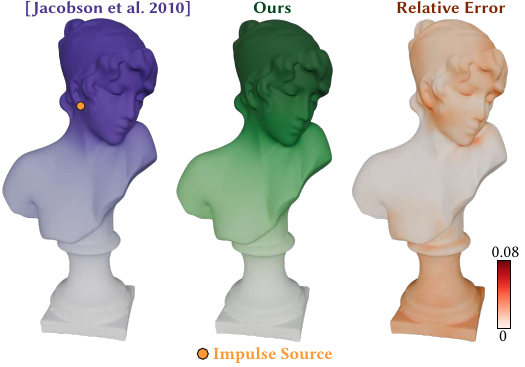}
  \caption{Green's functions of the biharmonic equation with Neumann boundary conditions. Our solution agrees closely with a ground truth obtained from mixed FEM operator of \citet{jacobson2010mixed}.}
  \Description{}
  \label{fig:biharmonic-scultpure}
\end{figure}